\documentclass[aps,prl,reprint,twocolumn,showpacs,superscriptaddress,amsmath,amssymb,citeautoscript]{revtex4-1}
\usepackage{amsmath}
\usepackage{amssymb}
\usepackage{graphicx}
\usepackage{color}
\usepackage{bm}
\usepackage{epstopdf}
\DeclareGraphicsExtensions{.eps}

\begin{document}

\title{Spontaneous Increase of Magnetic Flux and Chiral-Current Reversal in Bosonic Ladders: Swimming against the Tide}
\author{S. Greschner}
\affiliation{Institut f\"ur Theoretische Physik, Leibniz Universit\"at Hannover, 30167~Hannover, Germany} 
\author{M. Piraud} 
\author{F. Heidrich-Meisner}
\affiliation{Department of Physics and Arnold Sommerfeld Center for Theoretical Physics, Ludwig-Maximilians-Universit\"at M\"unchen, 80333 M\"unchen, Germany}
\author{I. P. McCulloch}
\affiliation{ARC Centre for Engineered Quantum Systems, School of Mathematics and Physics, The University of Queensland, St Lucia, QLD 4072, Australia}
\author{U. Schollw\"ock}
\affiliation{Department of Physics and Arnold Sommerfeld Center for Theoretical Physics, Ludwig-Maximilians-Universit\"at M\"unchen, 80333 M\"unchen, Germany}
\author{T. Vekua}
\affiliation{Institut f\"ur Theoretische Physik, Leibniz Universit\"at Hannover, 30167~Hannover, Germany} 


\begin{abstract}
The interplay between  spontaneous symmetry breaking in many-body systems, the wavelike nature of quantum particles and lattice effects produces an extraordinary behavior of the 
chiral current of bosonic particles in the presence of a uniform magnetic flux defined on a two-leg ladder. While non-interacting as well as strongly interacting particles, stirred by the magnetic field, circulate  along the system's boundary in the counterclockwise direction {in the ground state}, 
interactions { stabilize vortex lattices. These states break translational symmetry, which can lead to  a reversal of the  circulation direction}.
Our predictions could readily be accessed in quantum gas experiments with existing setups or in arrays of Josephson junctions.
\end{abstract}
\date{\today}

\maketitle

\begin{figure}[t]
\begin{center}
\includegraphics[width=.99\columnwidth]{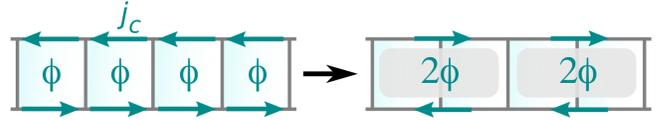}
\end{center}
\vspace{-0.6cm}
\caption{(Color online)   A spontaneous
 doubling of the unit cell  leads to an increase of the effective flux as well: $\phi \to 2\phi$. Under the conditions described in the text, the chiral current $j_c$ (indicated by  the arrows)  reverses its direction.\label{fig:sketch}}
\end{figure}

A charged {\it quantum particle}, due to its wavelike nature~\cite{Broglie},  picks up an 
increment to its phase proportional to the magnetic flux piercing the area enclosed by the particle's (closed) path. If the particle hops around a pla\-quet\-te, 
the accumulated phase is proportional to the magnetic flux $\Phi=B a^2$, $a$ being the  lattice constant. Since the phase of the wave-function is defined modulo $2\pi$, the
action of a magnetic field on quantum particles in a lattice is periodic: If we define a dimensionless flux $\phi= 2\pi \Phi/\Phi_0$ with the magnetic flux quantum $\Phi_0=h/q$, where $h$ 
is Planck's constant {and $q$ is the charge}, any physical quantity ${\mathcal A}$ obeys ${\mathcal A}(\phi)=  {\mathcal A}(\phi+2\pi)$. 
{The minimal extensive lattice on which this behavior can emerge is a two-leg ladder (see Fig.~\ref{fig:sketch}).}


In this work, we  discuss the intriguing effect of a {\it reversal} of the circulation direction of the chiral current of interacting bosons on a  
two-leg ladder,
due to the spontaneous formation of a large unit cell {\it without} changing the external magnetic field. The key ingredients to realize this effect are, first,
the wavelike nature of quantum particles defined on a lattice, and second, many-body effects.

The basic idea is sketched in Fig.~\ref{fig:sketch}. 
For a single plaquette, the magnetic fields corresponding to values of the flux  $0<\phi<\pi$ produce a ground-state net current with a counterclockwise chirality.
When one assembles these plaquettes into a minimal extensive lattice such as the  two-leg ladder, naively, 
one would expect that the local currents  on individual links along the boundary of the ladder add up to produce a 
net chiral current $j_c$, also circulating  in counterclockwise direction for $0<\phi<\pi$. Indeed, this is the case for non-interacting particles~\cite{Huegel}. 
In general, since the chiral current is a ground-state property, the
relevant unit cell does not need to be the same as the unit cell of the underlying lattice.

\begin{figure}[tb]
\begin{center}
\includegraphics[width=.99\columnwidth]{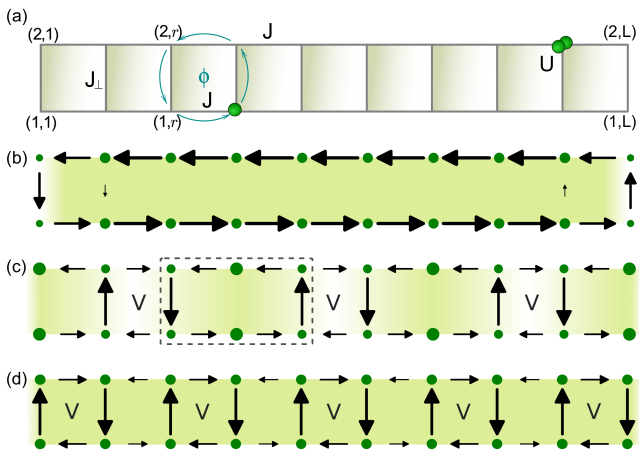}
\end{center}
\vspace{-0.6cm}
\caption{
(Color online) 
(a) Sketch: Two-leg ladder with $L$ rungs and a uniform flux $\phi$ per plaquette. 
(b)-(d) Local currents and onsite densities in (b) the Meissner phase, (c) the VL$_{1/3}$-SF and  (d) the VL$_{1/2}$-SF  [$U=2J$, $J_{\bot}=1.6J$, $\rho=0.8$ and (b) $\phi=0.6\pi$, $L=10$, (c)  $\phi=0.8\pi$, $L=120$   and  (d) $\phi=0.9\pi$, $L=120$]. 
The length of the arrows encodes the strength of local currents. The size of the circles and the intensity of the shading are proportional to the onsite density. 
In (c), a Meissner-like region is indicated by the dashed line. The chiral current is reversed in  (d), but not in (b) and (c). 
{The value of $J_\perp/J=1.6$ was chosen since two VLs are realized for these parameters}.
\label{fig:model}}
\end{figure}

If the unit cell is doubled, 
then this will result in a doubling of the effective flux  $\phi_{\rm eff} = 2 \phi$, piercing the enlarged unit cell of the ground state.
 The chiral current $j_c$ is an odd function of flux and periodic, thus $j_c(\phi)=j_c(\phi-2\pi)=-j_c(2\pi-\phi)$.  
As a consequence, for values of the flux $\pi/2<\phi<\pi$ defined with respect to the original cell, the
effective flux is in the domain $\phi_{\rm eff}\in (-\pi,0)$ modulo $2\pi$, which one would also obtain if the
orientation of the magnetic field was inverted. 
Here, we show that such an increase of the unit cell arises as a consequence of the spontaneous breaking of discrete lattice-translation symmetry
in vortex-lattice (VL) phases.

We study the single-band Bose-Hubbard model defined on a two-leg ladder lattice with flux $\phi$ per plaquette [see Fig.~\ref{fig:model}(a)]:
\begin{eqnarray}
H &=& -J \sum_{r=1 \atop \ell =1,2 }^{L-1} ( a^\dagger_{\ell,r+1} a_{\ell,r} + \textrm{H.c.} ) + \frac{U}{2} \sum_{r=1\atop \ell =1,2 }^{L} n_{\ell,r} (n_{\ell,r} -1)  \nonumber \\
   && -J_{\bot}  \sum_{r=1}^L ( e^{-i r \phi} a^\dagger_{1,r} a_{2,r} + \textrm{H.c.})\,. 
\label{eq:hamiltonian}
\end{eqnarray}
$a^\dagger_{\ell,r}$ creates a boson on the lower ($\ell=1$) or the upper site ($\ell=2$) of the $r$-th rung, $n_{\ell,r} = a^\dagger_{\ell,r}a_{\ell,r}$, and  $L$ is the number of rungs. The hopping matrix elements  between the nearest-neighbor sites along the ladder's legs and rungs are  denoted by $J$ and $J_{\bot}$, respectively, and $U$ is the repulsive
onsite interaction. The filling is $\rho=  N/(2L)$, where $N$ is the number of bosons.
We carried out  density matrix renormalization group (DMRG) simulations \cite{White,Uli} for $\rho < 2$ and $U \gtrsim J$.
For the complementary regime of large densities $\rho\gg 1$ and weak interactions $U\ll J \rho$, we use a mapping to a frustrated $XY$ model and apply a transfer-matrix approach~\cite{Kardar,Tang} (see the supplemental material \cite{suppmat} for  details on  both techniques).

Physical realizations are, for instance, either arrays of Josephson junctions of superconducting islands in a magnetic field~\cite{Ouden1,Ouden2,Roditchev}, where
the bosonic particles are Cooper pairs of electrons with $q=2e$, or neutral ultracold bosons in optical lattices \cite{Greiner,Dalibard} in the presence of
artificial gauge fields, subject to s-wave interactions~\cite{Bloch,Miyake,Aidel}. In the latter type of experiments, using either superlattices \cite{Bloch} or a synthetic lattice dimension \cite{celi,Spielman,Fallani}
to realize ladders, chiral currents have recently been measured.

\begin{figure}[b]
\begin{center}
\includegraphics[width=.99\columnwidth]{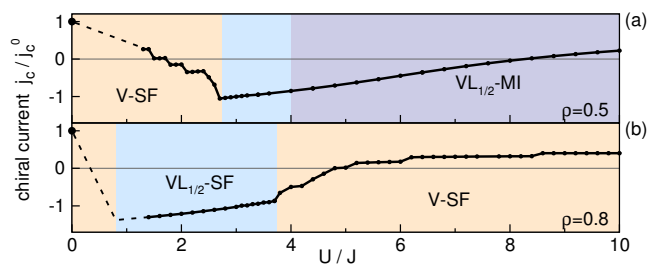}
\end{center}
\vspace{-0.6cm}
\caption{
(Color online) $j_c$ versus $U/J$ in the proximity of the VLs at $\rho_v=1/2$, for (a) $\rho=0.5$ and {(b) $\rho=0.8$,}
(DMRG data, $L=160$, $\phi=0.9\pi$, $J_{\bot}=1.6J$).
The exact value for $U=0$ is $j^0_c\simeq 0.08J>0$,  independently of $\rho$.
The small steps  seen in the VSF phase are finite-size effects (see  Sec.~S2 in \cite{suppmat}).
{The dashed lines serve as guide-to-the-eye for small values of U/J in the VSF phase, where for large system sizes (such as L=160), it is hard to converge DMRG with respect to both the local number of bosons  and the DMRG state space \cite{Uli, suppmat}. The dashed lines connect the $U=0$ value to the DMRG data
point for the smallest value of $U/J$ at which we have converged results (see Sec.~S6 \cite{suppmat}), plus we indicated the kink expected at the VL$_{1/2}$-SF to VSF boundary, based on  such data as shown in Fig.~S7 \cite{suppmat}.}
\label{fig:reversal}
}
\end{figure}

For the two-leg bosonic ladder, the existence of Meissner-like and vortex phases has been firmly established both in  the weakly- and strongly-interacting regime~\cite{Giamarchi,Petrescu,Mueller,Tokuno,us,petrescu2015}.
In Fig.~\ref{fig:model}(b), obtained from a  DMRG simulation, we depict the typical behavior of local currents and particle densities in the low-field Meissner phase. 
The currents are non-zero along 
the boundary of the two-leg ladder, consisting of the edge rungs and the legs of the ladder,
and vanish quickly on the inner rungs away from the first and last rung, producing a net chiral current  $ j_c = 
(\sum_{r} \langle j_{1,r}^{\parallel} -  j_{2,r}^{\parallel}\rangle + \langle j_{r=1}^{\perp} - j_{r=L}^{\perp}\rangle )/N$ (see  Sec.~S1 in \cite{suppmat} for definitions and properties of $j_c$). 
The coherence of the relative phase of bosons on the two legs  induces the Meissner phase in bosonic ladders~\cite{Kardar,Giamarchi} and hence a chiral current can also emerge in the Mott-insulating regime for fillings with an integer number of particles per rung~\cite{Petrescu,us}.  

Upon increasing the flux beyond a critical value, the system  
enters into a vortex phase where local currents on 
the inner rungs develop, resulting in a current configuration that is reminiscent of a Meissner phase disordered with vortices~\cite{Kardar,Giamarchi,us}. 
 These vortices interact repulsively with each other~\cite{Kardar}, yet for generic vortex densities $\rho_v$, they are distributed in the system without any periodicity. 
At certain commensurate vortex densities, VLs {were predicted to form} in the ground state \cite{Kardar,Giamarchi}. 
{DMRG results for  such VL states} are depicted in Figs.~\ref{fig:model}(c) and (d). 
In the VL superfluid~(SF) at vortex density $\rho_v=1/3$ [Fig.~\ref{fig:model}(c)], the currents on two neighboring pla\-quet\-tes form a complex 
[surrounded by a dashed line in Fig.~\ref{fig:model}(c)] that is a mini copy of the Meissner phase  (with 'screening' currents circulating around the complex' boundary and vanishing currents on its inner rung) such that on every third plaquette, a vortex resides,
around which the current circulates in the  direction opposite  to the behavior in the Meissner phase. 
In   Fig.~\ref{fig:model}(c), the chiral current  goes in the counterclockwise direction.
In the VL at $\rho_v=1/2$ (VL$_{1/2}$-SF), obtained for $\phi \lesssim \pi$,  a vortex sits on every other plaquette [Fig.~\ref{fig:model}(d)] and the direction of the chiral current is reversed.
{The observation of several stable VLs is a main result of this Letter.}

For not too low particle densities and flux values close to $\pi$, we generically observe a reversal of the chiral current tuned by the 
 interaction strength $U/J$: $j_c$ flows in a counterclockwise direction along the ladder's legs for $U=0$ and $U\to \infty$, yet for a certain range of interaction strengths at low temperatures the circulation direction of the current is reversed, as if the majority of particles {\it swam against the tide}.

The interaction driven chiral-current reversal is evident from the DMRG results presented in {Figs.~\ref{fig:reversal}(a) and (b)}. There, we plot the chiral current as a function of $U/J$ 
for $\rho=0.5$ and {$\rho=0.8$, respectively (see \cite{suppmat} for  data for $\rho=1$)}. 

We clearly numerically resolve a current reversal, which is tied to the presence of the VL at $\rho_v=1/2$. 
This state can be realized both in the superfluid (VL$_{1/2}$-SF) and the Mott-insulating regime (VL$_{1/2}$-MI).
The VL$_{1/2}$-SF phase is neighbored by a vortex superfluid (VSF) for small values of $U/J$. 
Generally, upon entering into the vortex-liquid phases from the VL phases, in the thermodynamic limit
the absolute value $|j_c|$ of the current decreases continuously (while for finite $L$,  small steps are seen~\cite{Orignac15}), showing a cusp-like behavior at the phase boundary
\cite{suppmat}. 
Other transitions that are crossed in Fig.~\ref{fig:reversal} do not leave a fingerprint in the interaction dependence of $j_c$
for $L\to \infty$.
 {For $\rho=0.5$}, further increasing $U/J$ takes the system from the VL$_{1/2}$-SF into a VL$_{1/2}$-MI.
 {Typically,} the sign of the chiral current becomes positive again
on the large $U/J$ side of the VL$_{1/2}$ phases. 
The chiral-current reversal is thus robust against the presence or absence of a mass gap and a variation of density. Moreover,
the absolute value of the reversed chiral current can exceed the $U=0$ value.

The spontaneous symmetry-breaking leading to the $M$-times enlarged unit cell with $\phi_{\rm eff}=M\phi \in (-\pi ,0)$ modulo $2\pi$ 
($M=2$ in Fig.~\ref{fig:reversal}) is a vital ingredient for obtaining the current reversal. 
We argue that yet another requirement is that the dominant contribution to the current comes from particles with a large wavelength experiencing the effective flux.
This is exemplified 
by the behavior of the chiral current in the fully gapped  VL$_{1/2}$-MI [see Fig.~\ref{fig:reversal}(a)]: 
Inside this state with a doubled unit cell, the chiral current, as a function of $U/J$, passes through zero and returns back to the original direction of rotation.
The localization length of bosons in the Mott-insulating phase becomes shorter with increasing $U/J$, restricting the typical wave-lengths of particles and as a consequence, less particles see the enlarged unit cell with doubled flux. On these grounds, we expect that in a given phase with a spontaneously enlarged unit cell, the absolute value of the reversed chiral current attains its maximum for the smallest $U/J$ in that phase, consistent with the data presented in Fig.~\ref{fig:reversal}. We emphasize that the optimal condition for observing the current reversal in two-leg ladders is a doubling of the unit cell.

{Interestingly, one could have deduced the existence of the chiral-current reversal from the flux-dependence of the ground-state energy shown in Ref.~\cite{JJL}
in the weak-coupling regime.  From  the bosonization analysis of~\cite{Giamarchi}, valid for $J_{\perp} \ll J$, one can 
also obtain the sign change. Yet, neither study actually
discussed the reversal effect.}
 DMRG results indicate that the chiral-current reversal driven by interactions also exists  on
 three-leg ladders~\cite{kolley}.

\begin{figure}[tb]
\begin{center}
\includegraphics[width=.99\columnwidth]{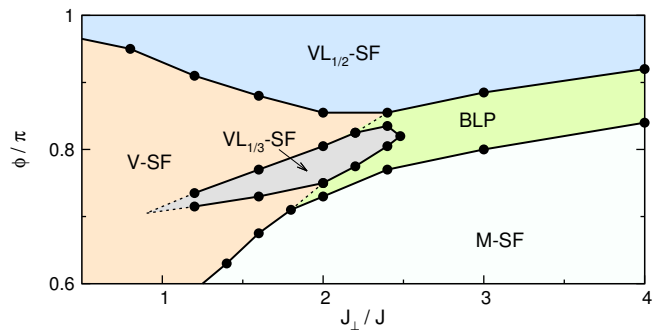}
\end{center}
\vspace{-0.6cm}
\caption{
(Color online) {{$\phi$ versus $J_\perp/J$ phase diagram at $U/J=2$ and $\rho=0.8$.}} {See the text for details.}
\label{fig:phasediag}
}
\end{figure}

The chiral current can also change its sign in fermionic ladders~\cite{carr,roux} as a function of  $\phi \in (0,\pi)$. {Yet, for fermions, the effect  
exists already in the 
noninteracting case as a result of the band structure
while for bosons,  the interaction-induced spontaneous flux increase is crucial for the current reversal.}

{To identify the regions in parameter space spanned by $\phi$, $U$, $\rho$ and $J_{\perp}$, in which the VL phases exist, we present a representative 
ground-state phase diagram in Fig.~\ref{fig:phasediag} for $U/J=2$ and $\rho=0.8$ as a function of flux and $J_\perp$ (see \cite{suppmat} for a density versus $U/J$ phase diagram). The figure illustrates the plethora of phases realized in the model:
the Meissner superfluid (MSF), a VSF, a VL$_{1/2}$-SF, a VL$_{1/3}$-SF and a biased ladder phase (BLP). The BLP, predicted in mean-field theory \cite{Mueller}, spontaneously breaks the symmetry between the two legs by imbalancing the density \cite{suppmat}. 
The VL$_{1/2}$-SF, which leads to the chiral-current reversal, is stable down to at least $J_{\perp} = J/2$.}

\begin{figure}[tb]
\begin{center}
\includegraphics[width=.99\columnwidth]{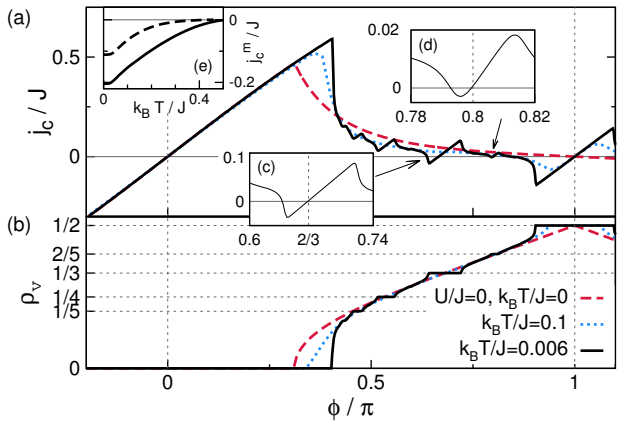}
\end{center}
\vspace{-0.6cm}
\caption{
(Color online)  
(a) Chiral current  $j_c/J$ and  (b) vortex density versus flux for $J_{\bot}=J/2$ {for  $\rho\gg 1$, $U\ll J\rho$}. 
{This value of $J_{\bot}$ is chosen since here, not only the VL at $\rho_v=1/2$ 
exhibits a chiral-current reversal at low $T$, but also those at $\rho_v=1/3$ and $2/5$~(see the insets (c) and (d)).}
The temperature is $k_BT=0.006J$ (continuous lines) and $k_BT=0.1J$ (dotted lines). The dashed lines are for $U=0$ at $T=0$. { (e) Largest negative value  $j_c^m = \mbox{min}_{\phi} j_c(\phi)$ 
of the reversed current for $J_\perp=0.5 J$ (dashed line) and $J_\perp=1.2 J$ (continuous line) as a function of $k_BT/J$.} {See \cite{suppmat}, Fig.~S6, for more data.}
\label{fig:temp}}
\end{figure}

 Since the current reversal is connected with spontaneous symmetry breaking in the ground state, 
 a relevant question pertains to the effect of temperature.
To obtain a quantitative understanding of $T>0$, we consider the limit of large particle densities $\rho \gg 1$ and weak interactions $U\ll J\rho$. 
Using the transfer-matrix approach~\cite{Tang}, we obtain the results shown in Fig.~\ref{fig:temp}, where the chiral current [Fig.~\ref{fig:temp}(a)] and vortex density [Fig.~\ref{fig:temp}(b)] versus flux are presented for different temperatures. Upon reducing temperature, VLs appear at $\rho_v=1/2,1/3,2/5,1/4,1/5,...$. These have an $M=2,3,5,4,5,...$-times enlarged unit cell, respectively~\cite{Kardar,Giamarchi,JJL,Tang}. At zero temperature, the vortex density, as a function of flux $\phi$, exhibits the famous devil's staircase structure \cite{Aubry,JJL} with a plateau for each commensurate value of $\rho_v$. The reversal of the current in the VL$_{1/2}$-SF phase clearly survives a finite temperature, which also applies (though for lower values of temperatures) to the current reversal in the vicinity of the VL$_{1/3}$-SF state for $\phi<2\pi/3$. {Thus, the reversal is more stable against thermal fluctuations than the underlying VLs}.
The optimal parameters for observing the current reversal in the weak-coupling regime are $\phi\simeq 0.9\pi$ and $J_{\bot}\simeq 1.2J$ where the reversal occurs for {$k_B T< J/2$ \cite{suppmat}}, {as is shown in Fig.~\ref{fig:temp}(e) as well as in Sec.~S5 in \cite{suppmat}}.

At extremely low temperatures and for the parameters of Fig.~\ref{fig:temp}, a reversal of the chiral current  is also visible in the $\rho_v=2/5$ VL state with $M=5$ for $\phi  \simeq 4\pi/5$ [see Fig.~\ref{fig:temp}(d)]. 
 Other VLs [corresponding to the regions in which  plateaus are formed in the $\rho_v=\rho_v(\phi)$ curve in Fig.~\ref{fig:temp}(b)] do not induce a reversal of the chiral current even at $T=0$ {since they are stable at values of fluxes for which $M\phi \in (0,\pi)$ modulo $2\pi$.}

A striking feature in the weak-coupling regime is the self-similar structure 
of the $j_c(\phi)$ curve for $k_BT\ll J$:  After the spontaneous $M$-fold increase of the unit cell, 
the chiral current exhibits a behavior that is similar to the one in the  Meissner phase. For instance, for the VL$_{1/2}$, 
this is visible in Fig.~\ref{fig:temp}(a)  at a flux value of $\phi=\pi$
and for VLs at  $\rho_v=1/3$ and $2/5$ [see Fig.~\ref{fig:temp}(c) and (d)] at $\phi=2\pi/3$ and $4\pi/5$, respectively. 
As a result, the absolute values of the reversed current for $\phi\sim \pi$ are several times larger than in the  $U=0$ case at the same value of $\phi$ 
(compare the continuous black curve and the red dashed curve in Fig.~\ref{fig:temp}(a) near $\phi=\pi$).

{We studied the behavior of interacting bosonic particles in the presence of a uniform magnetic flux on a two-leg ladder. 
Our work has three main results. First, we  determined  the range of stability of VLs predicted in \cite{Kardar,Giamarchi} 
and analyzed their microscopic structure. We  obtained representative phase diagrams, showing that the system is very rich, also realizing 
the BLP phase \cite{Mueller}. 
Second, we 
observed  the interaction-driven reversal of the chiral current  in the vicinity of certain VLs. 
Third, we  proposed an intuitive interpretation of the reversal via the spontaneous
increase of the effective flux   due to the enlarged unit cell in the VLs. We expect all these results to  influence experimental work on low-dimensional bosonic systems in the presence
of gauge fields. }

\begin{acknowledgments}
S.G. and T.V. acknowledge support by QUEST (Center for
Quantum Engineering and Space-Time Research) and DFG Research Training Group
(Graduiertenkolleg) 1729. We are grateful to N. Cooper, E. Jeckelmann,  A. L\"auchli, M. Lein, G. Roux,  and L. Santos for useful discussions {and
we thank T. Giamarchi for comments on an earlier version of the manuscript.}
The research of M.P. was supported by the European Union
through the Marie-Curie grant ’ToPOL’ (No.~624033)
(funded within FP7-MC-IEF). I.MC. acknowledges funding from the Australian
Research Council Centre of Excellence for Engineered Quantum Systems and grant number CE110001013.
\end{acknowledgments}

\bibliography{references}

\end{document}


\title{Supplemental  material for ``Spontaneous increase of magnetic flux and chiral-current reversal in bosonic ladders: Swimming against the tide''}
\author{S. Greschner}
\affiliation{Institut f\"ur Theoretische Physik, Leibniz Universit\"at Hannover, 30167~Hannover, Germany} 
\author{M. Piraud} 
\author{F. Heidrich-Meisner}
\affiliation{Department of Physics and Arnold Sommerfeld Center for Theoretical Physics, Ludwig-Maximilians-Universit\"at M\"unchen, 80333 M\"unchen, Germany}
\author{I. P. McCulloch}
\affiliation{ARC Centre for Engineered Quantum Systems, School of Mathematics and Physics, The University of Queensland, St Lucia, QLD 4072, Australia}
\author{U. Schollw\"ock}
\affiliation{Department of Physics and Arnold Sommerfeld Center for Theoretical Physics, Ludwig-Maximilians-Universit\"at M\"unchen, 80333 M\"unchen, Germany}
\author{T. Vekua}
\affiliation{Institut f\"ur Theoretische Physik, Leibniz Universit\"at Hannover, 30167~Hannover, Germany} 

\begin{abstract}
In this Supplemental Material we provide the  definition of the chiral current, discuss its finite-size effects 
and we show additional data for its dependence on the interaction strength.
We further discuss the phases  mentioned in the main text in more detail, elaborate on the  phase transitions and we 
show an additional phase diagram for density versus interaction strength.
 Additionally, we present patterns of the local current and density  used for the characterization of the ground-state phases. Finally, we describe 
in more detail the methods used to obtain the results of the main text.
\end{abstract}
\date{\today}

\maketitle

\setcounter{figure}{0}
\setcounter{equation}{0}
\renewcommand{\thetable}{S\arabic{table}}
\renewcommand{\thefigure}{S\arabic{figure}}
\renewcommand{\theequation}{S\arabic{equation}}
\renewcommand{\thesection}{S\arabic{section}}


\section{Chiral current}
\label{sec:jc}
The local current operators, defined through the standard lattice version of the continuity equation (setting the lattice constant $a=1$ and $\hbar=1$) for particle densities, 
are denoted by $j^{\parallel}_{\ell,r}$ (along the links of the ladder legs) and $j^{\bot}_{r}$ (on the rungs). Their explicit expressions are 
\begin{align}
j^{\parallel}_{\ell,r} &= i J \left( a^\dagger_{{\ell,r+1}} a_{\ell,r} - a^\dagger_{\ell,r} a_{{\ell,r+1}} \right)
\nonumber\\
j^{\bot}_{r} &= i J_{\bot} \left( e^{-i r \phi} a^\dagger_{1,r} a_{2,r} - e^{i r \phi} a^\dagger_{2,r} a_{1,r} \right)\,.
\label{eq:CurrentsDefs2}
\end{align}
In the thermodynamic limit ($L\to \infty$), where contributions from the  rungs at the boundaries ($r=1$ and $r=L$) of the ladder are negligible, or 
 for periodic boundary conditions, the chiral current becomes 
\begin{equation} 
j_c =\frac{1}{N}\sum_{r} \langle j^{\parallel}_{1,r}-j^{\parallel}_{2,r} \rangle =\frac{ \partial E_0}{\partial \phi},
\end{equation}  
where $E_0$ is the ground-state energy per particle~\cite{us} (in Secs. S1-S4 we will discuss ground-state properties (hence $T=0$)). 
 
One can easily show that for non-interacting bosons ($U=0$) and for  $J_{\bot}\le 2J\sin{\frac{\phi}{2}}\tan{\frac{\phi}{2}}$, 
\begin{equation}
j_c=\frac{J^2_{\bot} \sin{\phi}}{ 8J\sin^4{\frac{\phi}{2}}\sqrt{1+(J_{\bot}/2J\sin{\frac{\phi}{2}})^2 } }\,.
\end{equation}
For $U=0$, the chiral current saturates at $j_c=J\sin{\frac{\phi}{2}}$ in the Meissner-superfluid (MSF) phase, i.e., for $J_{\bot}>2J\sin{\frac{\phi}{2}}\tan{\frac{\phi}{2}}$.

In the complementary regime of strongly interacting bosons $U\gg J$, we also observe a quadratic dependence of $j_c$ on $J_\perp$ for small values 
of $J_{\perp}/J$. In fact, such a dependence holds for arbitrary values of $U/J$ for $J_{\bot}\ll J \phi$ ~\cite{us}. 
For strong  hopping on the rungs $J_{\perp}\gg J$ and for very strong interactions $U/J\to \infty$, the chiral current 
can decay to zero in the Mott-insulating states even at half-integer filling. For instance, for $\rho=0.5$, the chiral current is 
$j_c=J^2\sin{\phi}/(2J_{\bot})$~\cite{us}. Both in the $U=0$ and the $U/J\to \infty$ regime, the particles circulate counterclockwise along the boundary given that $0<\phi<\pi$.

\section{Chiral-current reversal and finite-size dependence of the chiral current}
\label{sec:fs}
\begin{figure}[tb]
\begin{center}
\includegraphics[width=1.0\columnwidth]{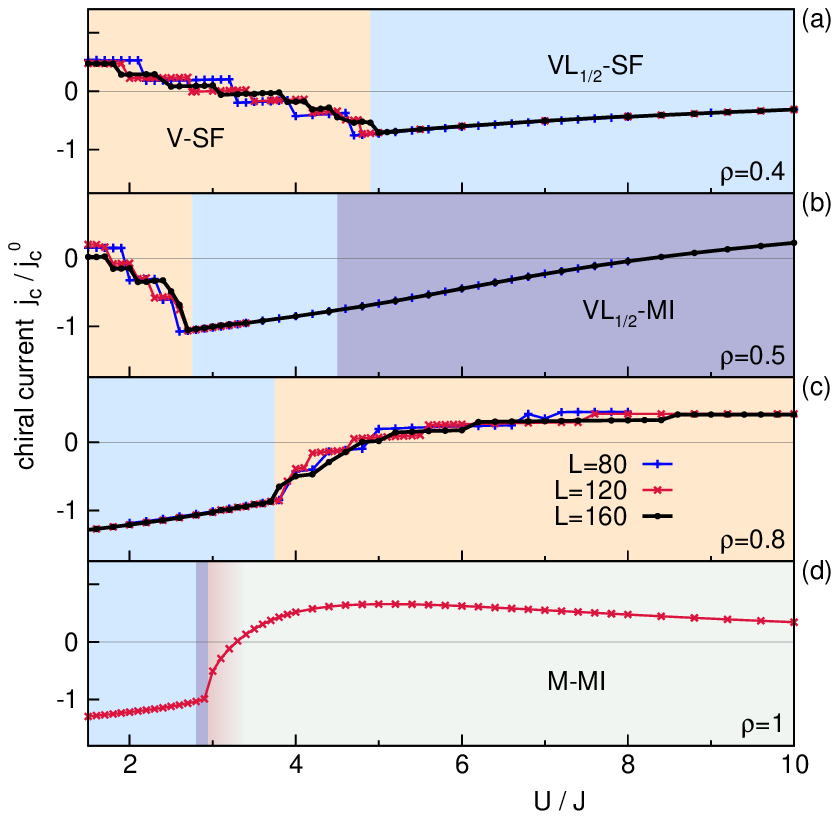}
\end{center}
\vspace{-0.6cm}
\caption{ {\it Finite-size dependence of the chiral current and different fillings}. We plot $j_c$ as a 
function of  interaction strength $U/J$ for the parameters of Fig.~3 of the main text ($\phi=0.9\pi$, $J_{\bot}=1.6 J$).
The figure shows results for three 
system sizes [$L=80$ (blue symbols), $L= 120$ (red symbols) and $L=160$ (black symbols)] and for particle densities  (a) $\rho=0.4$, (b)  $\rho=0.5$, (c) $\rho=0.8$ and (d) $\rho=1$. Due to the incommensurate vortex density a significant system-size dependence is present in vortex-liquid superfluid states (VSF), as opposed to the behavior in vortex-lattice states. For $\rho=1$ we just show points for $L=120$, which agree with the data for $L=80$ on the scale of the figure.
\label{fig:edgecurrent}}
\end{figure}

In Fig.~3 of the main text, we presented  results for the  chiral current for the largest system size ($L=160)$ used in our DMRG simulations.
Strong finite-size effects are present in the vortex-liquid phase (VSF). 
In Fig.~\ref{fig:edgecurrent}, we further illustrate these finite-size effects and the current reversal effect for different fillings by showing additional results for 
the  chiral current for systems with  $L=80$, $120$ and $160$ rungs. By contrast, in the 
vortex-lattice phases, finite-size effects are practically invisible on the scale of the figure.

As one can see in Fig.~\ref{fig:edgecurrent}, upon increasing system size, the width of the small quasi-plateaux of almost constant chiral current present in the VSF phases diminishes, indicating that in the thermodynamic limit, the  chiral current will become a continuous function of the flux in all phases as well as at the phase transitions between the vortex-liquid and the vortex-lattice phases presented in Fig.~3 of the main text. For a filling of $\rho=1$ (presented in Fig.~\ref{fig:edgecurrent}~(d)), the transition from the vortex-lattice state with increasing $U/J$ happens inside the Mott-insulating (MI) phase, with  finite-size effects being much less severe.

In the normalization adopted in Fig.~\ref{fig:edgecurrent}, the value of the chiral current is independent of $\rho$ and positive at $U=0$, implying that in all cases, the chiral current  undergoes a sign change in the vortex-liquid phases as the transitions to the vortex-lattice states  are approached.
The data shown in the figure further shows that the chiral current reversal occurs irrespective of density or whether the underlying state is superfluid (SF) or Mott-insulating.

\section{Ground-state phases}
\label{sec:ground}

\begin{figure}[tb]
\begin{center}
\includegraphics[width=0.99\linewidth]{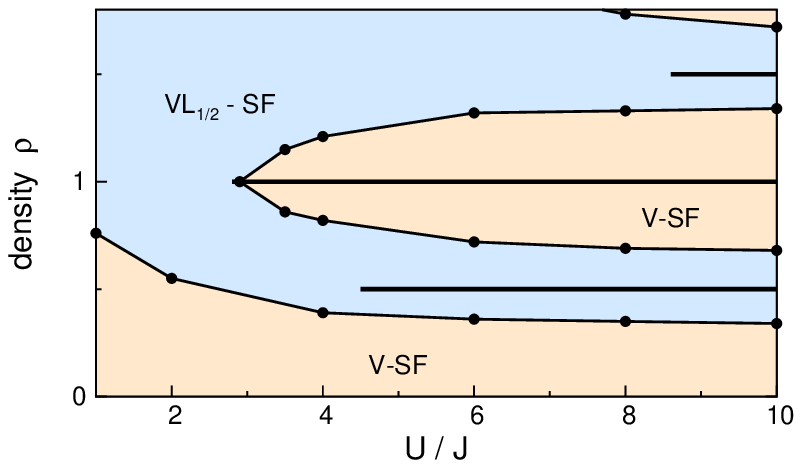}
\end{center}
\vspace{-0.6cm}
\caption{ {\it Ground-state phase diagrams}. Phase diagram in the  $(\rho,U)$ plane for $\phi=0.9\pi$ and $J_{\bot}=1.6J$.  The horizontal bold lines indicate Mott-insulating phases at commensurate fillings of $\rho=0.5,1,1.5$.
\label{fig:phasediag}}
\end{figure}

Next we discuss details of the phase diagrams in the parameter plane $(\phi$,$J_{\perp})$ for $U/J=2$  and a non-integer filling as shown in Fig.~4 of the main text. Additionally, we  present the phase
diagram in the  ($\rho, U/J$) plane for $\phi =0.9\pi$ and $J_{\perp}=1.6J$ in Fig.~\ref{fig:phasediag}. 

It is quite remarkable that the ground-state phase diagram shown in Fig.~4 of the main text
is so rich, with a plethora of superfluid phases.
Besides the MSF, the VSF, and the vortex lattice SF with vortex density $\rho_V=1/2$ (VL$_{1/2}$-SF), a VL$_{1/3}$-SF can get stabilized with a spontaneously tripled unit cell. However, the stability domain of the VL$_{1/3}$-SF [for the parameters of Fig.~4 of the main text] corresponds to flux values $\phi>2\pi/3$ and therefore, the increased (tripled) effective flux in the VL$_{1/3}$-SF state does not result in a current reversal, confirmed by our simulations. For larger values of $J_{\bot}/J$, an additional phase, predicted in mean-field approximation \cite{Mueller} (see also additional recent work \cite{tokuno15}), with a spontaneously emerging imbalance in the particle numbers on the two legs exists, dubbed biased ladder phase (BLP). This phase, however, preserves the original unit cell (see Fig.~\ref{fig:BLPcurrents}) and thus does not trigger a chiral-current reversal either.

In Fig.~\ref{fig:phasediag}, we present the phase diagram as a function of $\rho$ and $U/J$ for   $\phi=0.9\pi$, $J_{\perp}=1.6J$, $\rho< 2$ and $U\ge J$.
The vortex lattice with $\rho_v=1/2$ exists in a wide parameter range at sufficiently large densities (the critical lower density is $U$-dependent)
and, for larger $U/J \gtrsim 3$, is interlaced multiple times by the  VSF phase as density increases. Moreover, there are Mott-insulating phases at densities $\rho=1/2,1,3/2$ 
\cite{Petrescu,us,crepin11,keles}. The transition from the VL$_{1/2}$-SF into the VSF phase induced by changing $U$ or $\rho$ is caused by a proliferation of soliton-anti-soliton pairs in the VSF state (see Fig.~\ref{fig:localcurrents} in the next section).

The VSF is the only two-component Luttinger liquid~\cite{us} (central charge $c=2$), while 
all  other superfluid phases are one-component Luttinger liquids ($c=1$). The vortex-lattice and Meissner phases are identified by their characteristic local current configurations. They may be clearly discriminated from the vortex-liquid phase by calculating the central charge $c$ extracted 
from entanglement-scaling properties \cite{Calabrese04,Vidal03}, similar to the  strong-coupling $U/J\to \infty$ regime previously studied in \cite{us}.

All phase transitions in  Fig.~4 of the main text and Fig.~\ref{fig:phasediag} are second-order commensurate-incommensurate transitions~\cite{Giamarchi}, except for those into the BLP phase, which are identified by a sharply increasing particle density imbalance between the legs $\Delta n = \sum_r \left( n_{1,r}-n_{2,r}\right) / N$. 

There are also isolated Berezinskii-Kosterlitz-Thouless (BKT)~\cite{Berezinskii,Kosterlitz} critical points from Mott-insulating phases to superfluid states, along the paths of a constant particle density. 
Similarly the vortex lattice to vortex-liquid transitions for a constant vortex density are expected to be of BKT type.

 The Mott-insulating region at $\rho=1$ [shown also in Fig.~\ref{fig:edgecurrent}(d)] is richer and apart from
a narrow VL$_{1/2}$-MI region,  a Meissner Mott-insulator state (M-MI) state is realized with increasing $U/J$.
Our simulations on finite-size ladders suggest the possibility of  an intermediate vortex Mott-insulator (V-MI) region between the 
VL$_{1/2}$-MI and M-MI phases. It remains yet to be confirmed whether this intermediate V-MI survives in the thermodynamic limit or whether 
a direct transition between the VL$_{1/2}$-MI and M-MI phases is realized, similar to the case at $\phi=\pi$ \cite{dhar12,dhar13}.

\section{Local currents, particle densities and chiral current in different phases}
\label{sec:local}

\begin{figure*}[tb]
\begin{center}
\includegraphics[width=1.0\linewidth]{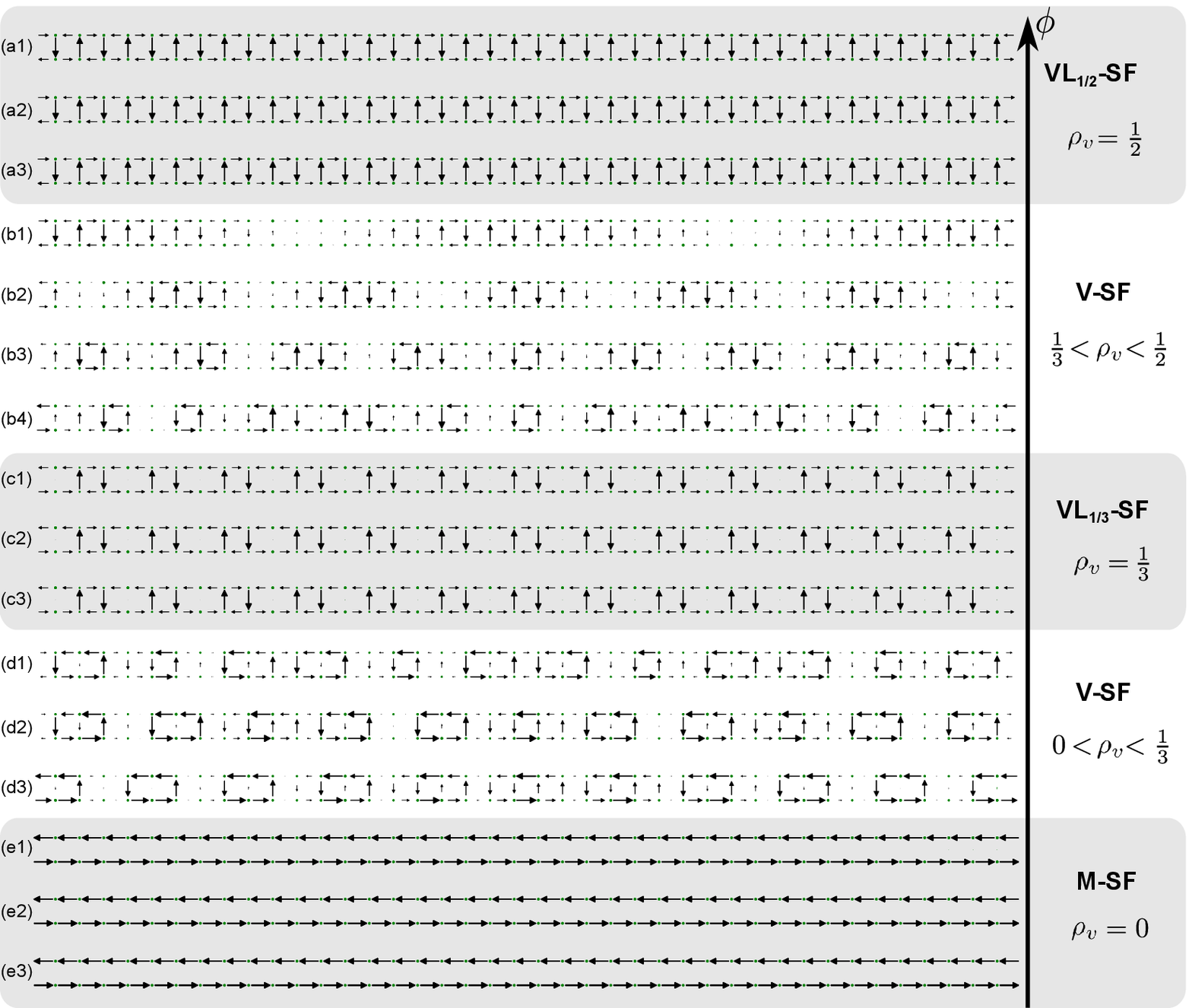}
\end{center}
\vspace{-0.6cm}
\caption{ {\it Configurations of local currents and densities in various phases}. DMRG results for the  cut at $J_{\bot}=1.6J$ through the phase diagram presented in 
Fig.~4 of the main textt  for $0.5\pi <  \phi <\pi$ ($U=2J$ and $\rho=0.8$). The configurations presented here correspond to the following flux values in descending order:  (a1) $0.95 \pi$,  (a1)   $0.93 \pi$, (a3)  $0.9 \pi$,  (b1)   $0.88 \pi$, (b2) $0.86 \pi$, (b3)  $0.82 \pi$, (b4) 
 $0.79 \pi$, (c1)  $0.77 \pi$, (c2)  $0.75 \pi$, (c3)  $0.73 \pi$, (d1)  $0.71 \pi$, (d2)  $0.69 \pi$, (d3)  $0.68 \pi$, (e1) 
 $0.64 \pi$, (e2)  $0.6 \pi$, and (e3)    $0.56 \pi$  ($L=120$ rungs). We only depict the behavior for the middle rungs of the ladder ($40 \le  r \le 80$) in the bulk of the system. The local currents and local particle densities presented in Fig.~2(c)-(d) of the main text in the VL$_{1/3}$-SF and VL$_{1/2}$-SF are cut from the central region of configurations (c2) and (a3) presented here. The length of the arrows encodes the absolute value of 
 the local currents and the size of the  circles encodes the  onsite density. In the top four configurations (a1)-(b1), the chiral current is reversed. 
\label{fig:localcurrents}}
\end{figure*}

{\it Currents and densities at $U=2J$, $\rho=0.8$.}
The different superfluid phases discussed in the main text can be distinguished from each other by the 
structure of local particle currents and local particle densities. 
In Fig.~\ref{fig:localcurrents}, we present typical examples for 
a cut through the phase diagram of Fig.~4 of the main text at $J_{\bot}=1.6J$. 
We show results  for a central portion of the ladder with $L=120$ rungs.

 Upon increasing $\phi$ (from bottom to top in Fig.~\ref{fig:localcurrents}), the system passes from the Meissner phase [MSF, Figs.~\ref{fig:localcurrents}(e1)-(e3)] (stable for $ \phi \lesssim
 0.7\pi $) to the  vortex-lattice state  VL$_{1/2}$-SF 
(shown in Figs.~\ref{fig:localcurrents}(a1)-(a3), realized for $\phi\lesssim \pi$) by going through 
additional intermediate vortex-liquid and vortex-lattice phases. As already mentioned, in the Meissner phase, currents are non-zero 
only along the ladder's legs (and, for open boundary conditions, also on the boundary rungs $r=1$ and $r=L$, not shown), hence the vortex density $\rho_v$ vanishes, while the 
 particle density is uniform in the bulk of the ladder. Adjacent to the Meissner phase, there is a vortex-liquid superfluid (VSF) with a 
finite vortex density $0<\rho_v<1/3$ [shown in Figs.~\ref{fig:localcurrents}(d1)-(d3)]. 
For values of  $ 0.75 \pi  \lesssim \phi \lesssim 0.82\pi $, the vortex-lattice phase with $\rho_v=1/3$  (VL$_{1/3}$-SF) becomes the ground state  
[shown in Figs.~\ref{fig:localcurrents}(c1)-(c3)].
Both local currents along the legs and along the rungs have a periodicity with a period of three times the lattice spacing, which is also exhibited by  
the local particle density. 

Upon further increasing the flux but still before the vortex lattice at $\rho_v=1/2$ (VL$_{1/2}$-SF) is stabilized, the system reenters into the vortex-liquid phase with vortex densities $1/3<\rho_v<1/2$ [shown in Figs.~\ref{fig:localcurrents}(b1)-(b4)]. In the vortex-liquid state and in the proximity of the VL$_{1/2}$-SF, 
domain walls (or solitons) between the two degenerate ground states of the VL$_{1/2}$-SF state, which we denote by $|1\rangle$ and $|2\rangle$, are clearly visible [see 
Fig.~\ref{fig:localcurrents}(b1) and (b2)]. We call a soliton a defect in the periodic structure of the VL$_{1/2}$-SF state with a $|1\rangle$-like pattern to  the left of the defect and a  $|2\rangle$-like pattern to the  right side and anti-solitons a defect with a $|2\rangle$-type pattern on the left side and a $|1\rangle$-type pattern on the right side. 
The number of solitons and anti-solitons in the vortex-liquid state in the vicinity of the VL$_{1/2}$-SF are identical, due to the topological confinement stemming from the double degeneracy of the ground state of the VL$_{1/2}$-SF phase. The local density shows modulations near the positions of solitons and anti-solitons. In the VL$_{1/2}$-SF state, on the 
contrary, the local densities are constant in the bulk along the ladder and the local currents show a modulation with a period of two lattice cells. 
In general, in vortex-liquid phases, the  particle density and local currents exhibit a complicated structure, unlike the highly regular patterns of the  vortex-lattice states.

The maximal possible vortex density in this system can in fact not exceed $\rho_v=1/2$.
This limit is reached for values of the flux $\phi\sim \pi$, in agreement with the picture originating from the weak-coupling regime $\rho\gg 1$,~\cite{Tang,JJL}.
The current configurations presented in our VL$_{1/3}$-SF and VL$_{1/2}$-SF phases, here obtained  from our numerical simulations and shown in 
Figs.~\ref{fig:localcurrents}(c1)-(c3) and  Figs.~\ref{fig:localcurrents}(a1)-(a3), respectively, are different from those depicted in~\cite{Giamarchi,Petrescu}. 
Namely, and most importantly, around the vortices, particles circulate in the direction opposite to the behavior in the Meissner phase. 
Hence, the vortices that we observe resemble the vortices of type-II superconductors. We further note that in VL$_{1/3}$-SF vortex lattices, 
the  particle density is reduced at the corners of those plaquettes at which the  vortices  sit,
 which is not captured in the standard linearized bosonization approach used in ~\cite{Giamarchi,Tokuno,Petrescu,us}.
By contrast, in the  VL$_{1/2}$-SF, vortices occupy every other plaquette and hence there is no corresponding effect on the local particle density. The density pattern of 
the VL$_{1/3}$-SF state is, however, inhomogeneous, yet nevertheless follows the periodicity of the local current configuration of the underlying vortex-lattice state: Vortices sit in regions of lower particle density. 
In the VL$_{1/3}$-SF, a periodically modulated local current and a periodic density modulation coexists with superfluidity.

Unlike linearized bosonization, a mean-field approach~\cite{Mueller} yields a modulation of the particle density in vortex phases.  
However, in mean-field theory and at $\phi=\pi$, the particle density is modulated from rung to rung with a period of two lattice constants, whereas our numerical simulations for fluxes close to $\pi$ (including $\phi=\pi$) produce VL$_{1/2}$-SF states with a uniform particle density along the ladder. 

\begin{figure}[tb]
\begin{center}
\includegraphics[width=.99\columnwidth]{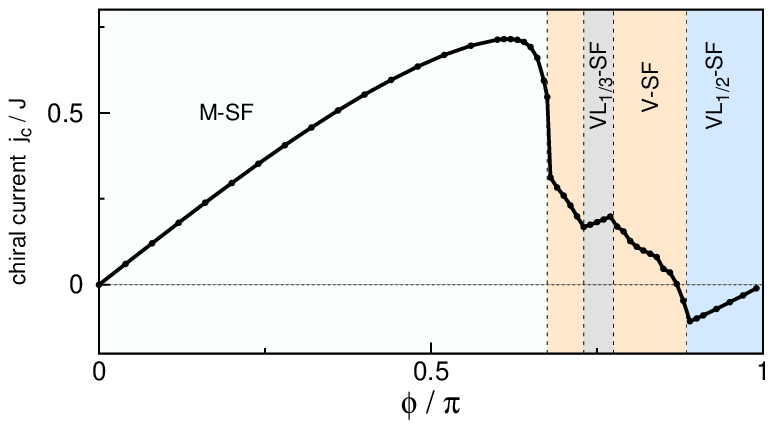}
\end{center}
\vspace{-0.6cm}
\caption{{\it Flux dependence of the chiral  current.} We display DMRG results for the chiral current along the cut at $J_{\bot}=1.6J$ through Fig.~4 of the main text  ($U=2J$ and $\rho =0.8$) for $0 \le  \phi \le \pi$, obtained for $L=120$ rungs. \label{fig:ec}}
\end{figure}

{\it Chiral current as a function of flux.} The dependence of  the chiral current as a function of flux $\phi$ and as the system goes through the different ground-state configurations shown in Fig.~\ref{fig:localcurrents} is presented in Fig.~\ref{fig:ec}. For $J_{\bot}=1.6J$, the chiral current does not undergo a sign reversal in the VL$_{1/3}$-SF state, since this state is stabilized only for $\phi>2\pi/3$. 
A reversal occurs only near $\phi\sim  \pi$. 
Even though generically, the  transition from Meissner to vortex-liquid phases is a second-order commensurate-incommensurate transition, in the numerical data for a finite system size shown
in Fig.~\ref{fig:ec}, there is  a finite jump in the chiral current. This jump, observed in our finite-size simulation, can be  attributed to the proximity of the multicritical point at which the Meissner superfluid phase, the vortex superfluid and the BLP phase meet (see Fig.~4 of the main text).

\begin{figure}[tb]
\begin{center}
\includegraphics[width=.99\columnwidth]{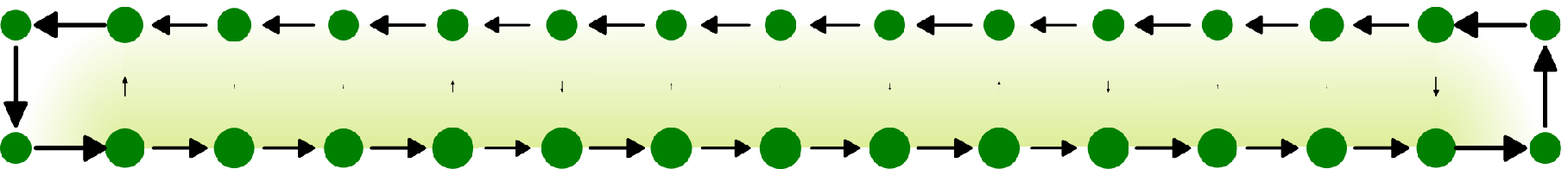}
\end{center}
\vspace{-0.6cm}
\caption{{\it Typical pattern of local currents and local densities in the biased ladder phase (BLP)}.
 In the particular realization shown here the lower leg hosts the 
majority of particles. These DMRG results  are for the parameters of Fig.~4 of the main text  ($\rho=0.8$ and $U=2J$) and for a point inside the 
BLP phase ($\phi=0.8\pi$, $J_{\bot}=2.7J$, $L=15$).  The length of the arrows encodes  the strength  of local currents and the size of the 
circles  is proportional to the  density.\label{fig:BLPcurrents}}
\end{figure}

{\it Local currents and particle densities in the BLP phase.} For completeness we also present results for the typical behavior of local currents and particle densities in the 
BLP phase, obtained from DMRG simulations for $L=15$, shown in Fig.~\ref{fig:BLPcurrents} (we choose a small system to illustrate the 
 behavior on the boundary rungs $r=1$ and $r=L$ as well). 
The local currents are non-zero along the legs and edge rungs of the ladder and particles flow in
counterclockwise direction. The absolute values of local currents (which for a finite system depend on the distance from the edge rungs) are identical in both legs, while the  density is imbalanced between the two legs.
Thus,  one can conclude  that the behavior of the chiral current in the  BLP superfluid is quite similar to that in the Meissner superfluid,
where particles circulate along the ladder's boundary, vanishing quickly on the  rungs in the bulk since $\rho_v=0$).
Nevertheless, the BLP phase is distinct from the MSF since in the former there  is a spontaneously broken discrete $Z_2$ symmetry corresponding to time reversal combined with exchange of the legs.

\section{Temperature dependence in the weak-coupling limit} 
\begin{figure}[b]
\begin{center}
\includegraphics[width=.99\columnwidth]{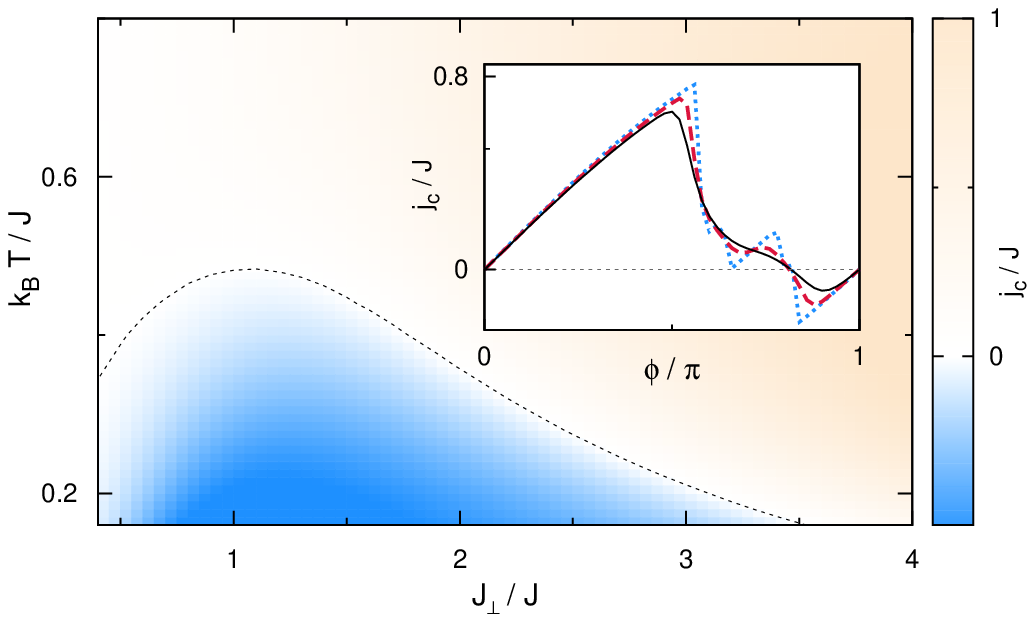}
\end{center}
\vspace{-0.6cm}
\caption{{\it Temperature dependence of the chiral current in the weak-coupling limit}. The coloring  in the main panel encodes the strength of the chiral current $j_c/J$ as a function of temperature $k_B T /J$ and rung-coupling $J_\perp/J$ for $\phi=0.92\pi$. The curve defined as $j_c/J=0$ is highlighted with a dotted line. The inset shows $j_c/J$ for $J_\perp/J=1$ as a function of the flux $\phi$ for several temperatures $k_B T/J=0.01$ (dotted blue line), $0.1$ (dashed red line) and $0.2$ (solid black line). \label{fig:wcfiniteT}}
\end{figure}

In Fig.~\ref{fig:wcfiniteT}, we present additional results on the temperature dependence of the chiral-current reversal. As mentioned in the main text one can observe the current reversal related to the presence of a VL$_{1/2}$ state close to $\phi=\pi$ for   $k_B T < J/2$. The inset of Fig.~\ref{fig:wcfiniteT} shows the $\phi$-dependence of $j_c$ for several temperatures for $J_\perp=J$. For the other vortex-lattice phases the current-reversal effect is  washed out by increasing temperature faster.
\section{Methods}
\label{sec:methods}

\begin{figure*}[t]
\begin{center}
\includegraphics[width=.99\columnwidth]{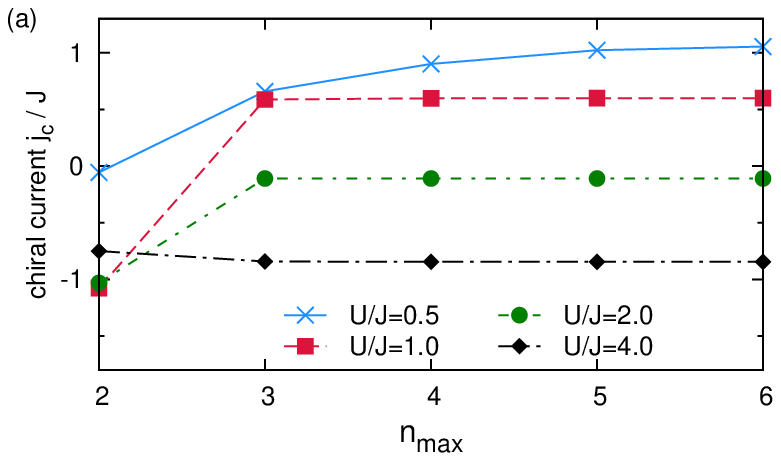}
\includegraphics[width=.99\columnwidth]{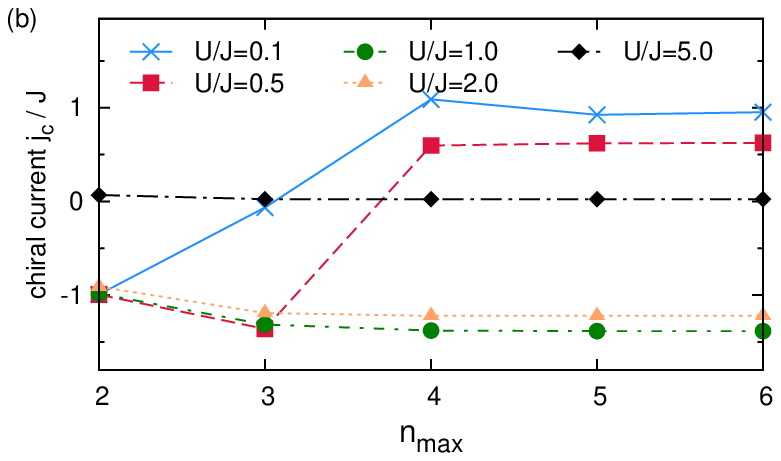}
\end{center}
\vspace{-0.6cm}
\caption{{\it Dependence of the chiral current on the bosonic cutoff $n_{max}$}. DMRG calculation for the parameters of Fig.~3, $\phi=0.9\pi$ and $L=40$ rungs for density (a) $\rho=0.5$, and (b) $\rho=0.8$.
\label{fig:jc_nmax}}
\end{figure*}

 {\it DMRG.} For interaction strengths of $U/J> 1$ and particle filling $\rho<2$, we perform large
scale numerical density matrix renormalization group 
\cite{White, Uli} simulations which allow us to calculate zero-temperature
properties for open boundary conditions and systems with up to $L=160$ rungs, using 1000 DMRG states. 
Since the local Hilbert space of bosons is generally unconstrained we 
employ a cutoff of the maximal occupation.
We typically use a cutoff of four bosons
per site which is justified due to the repulsive nature of the onsite interactions and for 
$U>J$. By comparison with larger and smaller cutoffs we have verified the numerical accuracy of the quantities shown here. 
Figure~\ref{fig:jc_nmax} shows the dependence of the chiral current on the bosonic cutoff $n_{\rm max}$. For $U\gtrsim J$, the results, obtained by keeping not more than $n_{\rm max}=4$ particles per site, are already well converged. For $0<U\lesssim J$ and by further  increasing $n_{\rm max}$, we have verified that the sign of the chiral current becomes positive again 
in the VSF phase for some $U<U_c<J$. In Fig.~3 of the main text we indicate this trend, based on data such as the ones shown in Fig.~\ref{fig:jc_nmax}, with a dashed line connecting DMRG data of $U\sim 1.5 J$ and the known value at $U=0$.
Moreover, the dashed lines in Fig.~3 of the main text
indicate the kink that we expect to be present at the
VL$_{1/2}$-SF to VSF boundary, based on the data of Fig.~\ref{fig:jc_nmax}.

Close to the VSF to VL$_{\rho_v}$-SF boundaries, the DMRG simulations tend to converge
to metastable excited states with larger or smaller vortex density. Here,
we have performed several calculations starting from different randomly chosen
initial states. Selecting the lowest energy states gives the piecewise
continuous chiral-current curve presented in Fig.~\ref{fig:edgecurrent} and Fig.~3 of the main text.

 The BKT transitions between VL$_{1/2}$-SF and VL$_{1/2}$-MI 
at fillings $\rho=0.5$ and $\rho=1.5$ shown in Fig.~\ref{fig:phasediag} and Fig.~3~(a) of the main text are determined by numerically analyzing correlation functions. At the transition from the 
VL$_{1/2}$-SF to the VL$_{1/2}$-MI, bosonization predicts that the single-particle correlation function decays as $ \langle a^{\dagger}_{l,r}a_{l,r+x} \rangle  \sim x^{-\frac{1}{4}}$ (up to logarithmic corrections). The transition point can be extracted accurately from the finite-size scaling behavior of peaks in the quasi-momentum distribution function~\cite{greschner13}.

For the BKT transition from the superfluid to the Mott insulator at unit filling  $\rho=1$, which 
happens inside the vortex lattice at $\rho_v={1/2}$ state presented in Fig.~\ref{fig:phasediag} and Fig.~\ref{fig:edgecurrent}~(d), 
it can be shown using bosonization that the single-particle correlation function decays as $ \langle a^{\dagger}_{l,r}a_{l,r+x} \rangle  \sim x^{-\frac{1}{8}}$ (up to logarithmic corrections). Other phase transitions are estimated from the behavior of the chiral current as well as from the local-current structure (such as the transitions  indicated in Fig.~\ref{fig:edgecurrent} and Figs. 3 of the main text).

{\it Transfer-matrix approach.}
For large particle densities per ladder site $\rho\gg 1$, we follow an approach used in the theory of Josephson-junction ladders, where each superconducting island is proximity-coupled to its three neighbors.
It is convenient to introduce a density phase representation of bosonic operators $a^{\dagger}_{l,r}=\sqrt{\rho+\delta \rho_{l,r}}e^{i\theta_{l,r}}$, where $\delta \rho_{l,r}=n_{l,r}-\rho$ and $\theta_{l,r}$ describe the density fluctuations and the phase of the condensate at site $(l,r)$, respectively. For $U\ll J\rho$, neglecting the charging effects of the superconducting islands~\cite{Tang,JJL}, the Hamiltonian of Eq.~(1) in the main text becomes  
\begin{eqnarray}
H& \to & -2J\rho \sum_{\ell =1,2;r=1}^L  \cos( \theta_{\ell,r+1} -\theta_{\ell,r}) \nonumber\\
&&-2J_{\bot} \rho \sum_{r=1}^L \cos( \theta_{1,r}-\theta_{2,r} -r\phi), \label{eq:class} 
\label{eq:hamiltonianJJL}
\end{eqnarray}
 with local variables $\theta_{l,r}\in [0,2\pi)$.
In Eq.~\eqref{eq:class}, it is more convenient to impose periodic boundary conditions.
The transfer matrix $\hat P$, connected with the partition function $Z=\text{tr\, }\hat P^L $ was obtained in~\cite{Tang}.
Following \cite{Tang}, we numerically calculated the eigenvalues $\lambda_n$ of the transfer matrix 
(where $n=0,1,2...$ and $|\lambda_n|\ge |\lambda_{n+1}|$) in the thermodynamic limit. The periodicity of the argument of the second largest eigenvalue as a function of  
flux  yields the vortex density  $\rho_v(\phi)= \mbox{Arg}[\lambda_1(\phi)]/2\pi$ for different temperatures~\cite{Tang} presented in Fig.~5~(b) in the main text. The chiral-current curves corresponding to different temperatures, presented in Fig.~5~(a) in the main text, are obtained from 
$$j_c(\phi)=-\frac{k_BT}{N}\frac{\partial {\ln Z}}{\partial \phi }\,.$$

\bibliography{references}